# Tailoring Infrared Absorption and Thermal Emission with Ultrathin-film Interferences in Epsilon-Near-Zero Media

*Ben Johns\*, Shashwata Chattopadhyay and Joy Mitra\**

School of Physics, Indian Institute of Science Education and Research, Thiruvananthapuram, India 695551

E-mail: ben16@iisertvm.ac.in, j.mitra@iisertvm.ac.in

**Abstract**: Engineering nanophotonic mode dispersions in ultrathin, planar structures enables significant control over infrared perfect absorption (PA) and thermal emission characteristics. Here, using simulations, the wavelength and angular ranges over which ultrathin, low loss, epsilon-near-zero (ENZ) films on a reflecting surface most efficiently absorb and re-radiate are identified, and the design parameters that tailor the ENZ mode dispersion within these limits are investigated. While the absorption is spectrally limited to wavelengths where the refractive index ($n$) lies below unity, the angular limits are determined by the ENZ material dispersion in this range. A model of ultrathin-film interference is developed to provide physical insight into the absorption resonances in this regime, occurring well below the conventional quarter-wavelength thickness limit. Driven by non-trivial phase shifts incurred on reflection at the $n < 1$ surface, these resonant interferences are shown to be universal wave phenomena in planar structures having appropriate index contrast, extending beyond ENZ materials. Selective choice of material, film thickness and loss allows fine-tailoring the mode dispersions, enabling wide variation in spectral range (∼ 0.1 – 1.0 $\mu$m) and precise directional control of spectrally and angularly narrow-band PA and thermal radiation, paving the way towards efficient ENZ-based infrared optical and thermal coatings.

# 1. Introduction

Nanophotonics, with its ability to control and manipulate light at the nanoscale, has led to the realization of various novel applications such as sub-diffraction waveguiding,[1] imaging,[2] optical cloaking [3] and single molecule sensing.[4] These unusual optical responses are often derived from the sub-wavelength contrast in optical properties of nanophotonic structures such as metasurfaces [5] and nanoparticle arrays.[6] However, recent advances have shown that optically thin films in planar structures can achieve substantial control in modifying the reflected and transmitted light spectrum exploiting interface phase shifts that differ significantly from 0 or $\pi$[7-10], which can be tuned to a desired value by engineering parameters like film thickness, doping etc. The absence of nanostructuring, large area scalability, and reduced fabrication and material costs make such ultrathin films ($d << \lambda$, $d$ is the film thickness and $\lambda$ is the wavelength of light) attractive for applications in coloring, filtering, enhanced absorption in photovoltaic applications, infrared absorbers and emitters, and in reconfigurable flat optics.[11-20]

One class of thin film media that has gained much interest recently are epsilon-near-zero (ENZ) materials, [21,22] which have a vanishing real part of permittivity ($\epsilon'$) at a specific frequency. ENZ materials offer strong light-matter interactions and show exciting effects such as super-coupling,[23] nanoscale field confinement and enhancement,[24,25] radiation control[26,27] and extreme non-linearities.[28] Further, ENZ thin films are known to support radiative modes called Berreman modes, which have a nearly flat dispersion near the frequency where $\epsilon' \to 0$.[29-31] Perfect coupling of *p*-polarized free space light into these radiative modes has been exploited to design efficient polarization switching,[32] high harmonic generation,[33] and tunable and broadband perfect absorption (PA).[34] This strong absorption in ENZ thin films, originating from the large field enhancements imposed by field continuity and the vanishing permittivity,[25] also results in a high thermal emissivity whose spectral and angular characteristics are determined by the dispersion of the radiative modes.[35] Therefore, by tailoring the modal dispersion, both the spectral and angular response of thermal emission can be controlled, [35,36] varying from spectrally selective, narrow[37] or wide-angle emission,[38,39] to broadband and directional emission.[40] The building block of these structures, the ENZ-metal bilayer, is known to support PA for *p*-polarized light, which has been explained by various effects such as coherent cancelling,[41] effective impedance matching [42,43] and critical coupling to ENZ modes.[34,44] However, various properties crucial to the engineering of thermal radiation sources remain to be investigated such as the available ranges, both spectral and angular, of thermal emission from an ENZ thin film, which are important in deciding the overall characteristics of a multilayer emitter.[40] Further, the degree to which the dispersion can be engineered by systematically varying the ENZ thickness and controlling the loss has yet to be elucidated, which will determine the fine-tunability of its thermal emission features.[45,46] It is therefore important to obtain a physical picture that not only provides a quantitative insight into the absorption and dispersion in ENZ layers, but also identifies their readily tunable parameters for tailoring the response.

Here, we show that the angular and spectral range of electromagnetic absorption and thermal radiation in ENZ thin films is restricted to a characteristic regime determined by the dielectric function of the ENZ material relative to the ambient superstrate medium. Using doped cadmium oxide (CdO) and $SiO_2$ ENZ layers on a reflecting (metallic) substrate with ambient air superstrate as model systems in the near-IR and mid-IR range respectively, we numerically demonstrate that perfect absorption of light in ultrathin layers is restricted to the spectral range where the ENZ layer is a dielectric ($\epsilon' > 0$) with refractive index (*n*) below unity. Further, the radiative modes giving rise to the absorption features are observed to be restricted to angles above a wavelength-dependent critical angle $\theta_c$. For practical purposes, the $\theta_c$ obtained is equivalent to that defined for total external reflection (TER) at the air-ENZ interface, for the corresponding lossless ENZ material. Indeed, a critical angle is undefined for lossy media, especially for ENZ media at regimes where $\epsilon' \sim 0$. Remarkably, the observed trends in ultrathin film absorption are not merely limited to ENZ materials, but are shown to be universal features of structures having an appropriate index contrast between the ambient medium ($n_1$) and thin film ($n_2 + ik_2$), such that $n_1 > n_2$. This indicates that the driving mechanism of absorption in these systems, generally described in terms of the Berreman mode excitation, can be explained as a more general wave interference phenomenon applicable even in media that do not support polaritonic resonances. To explain these unique features, we develop a model of thin film interference in $n < 1$ ultrathin films, which predicts that these films can sustain resonant interferences well below the quarter-wave thickness limit. The key to this phenomenon is the external reflection (ER) effect at the air–ENZ interface [47-49] arising from the refractive index contrast, due to the simultaneous realization of below unity refractive index and dielectric nature ($\epsilon' > 0$) of the ENZ layer. The angle-dependent ER phase shifts provide the required phases to satisfy the conditions for destructive interference of reflected light, making it tolerant to a fixed reflection phase due to the reflecting substrate or the near-zero propagation phase in the ultrathin film. Through straightforward relations connecting the ER phase and magnitude at the interfaces, we explain the spectral and angular confinement of the resonant modes and show how the dispersions and the spectral and angular characteristics of the resulting thermal emission can be finely tailored by choosing appropriate ENZ layer thicknesses and loss. Finally, we show that PA in the *low-loss* limit in $n < 1$ media is achieved only in the vicinity of the ENZ regime owing to the strong field enhancements therein, removing the necessity of high material loss in the ultrathin absorbers and emitters. The general wave interference mechanism of ultrathin film absorption elucidated here will be useful in the design of thermal emitters by providing design principles to engineer the dispersion in sub-wavelength modes and expanding the choice of available materials. Moreover, the identified characteristic spectral and angular ranges of ENZ materials will be useful in controlling thermal radiation in multilayer ENZ films or metasurfaces for spectrally- or angularly- selective infrared emission and absorption.

## 2. Results and discussion

### 2.1. Mechanism of ultrathin film interference

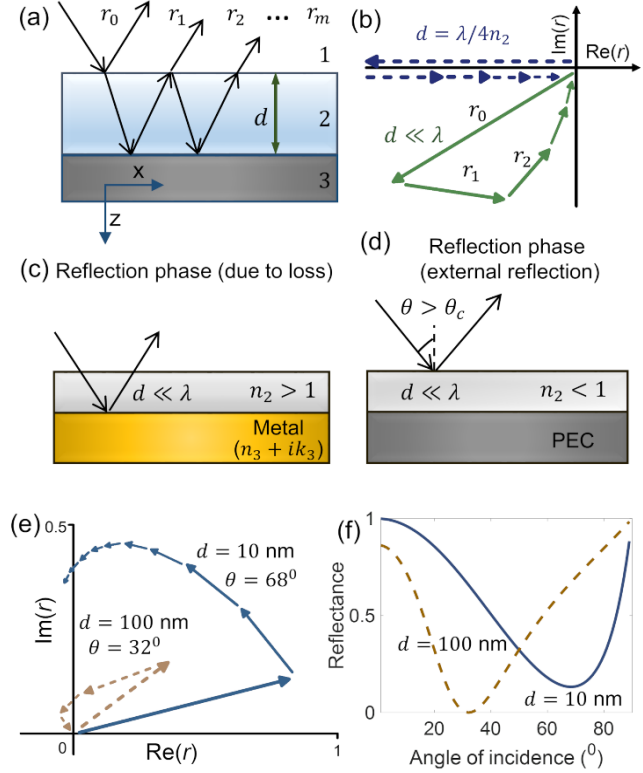

**Figure 1**. (a) Schematic of light incident on a thin film on an opaque substrate and resulting partial waves $r_m$. (b) Phasor diagram of partial waves in a PEC-backed, quarter-wave resonator ($d = \lambda/4n_2$, dashed arrows) and an absorbing-metal-backed, ultrathin resonator ($d << \lambda$, solid arrows) (c) When the opaque substrate is a weak reflector, phase shifts due to loss at the film-substrate interface allow resonances for $d << \lambda$. (d) Resonant absorption in a PEC-backed ultrathin resonator requires non-trivial phase shifts at the air-film interface, which is achieved when $n_2 < 1$. (e) Phasor diagrams for the case in (d) with $n_2 = 0.45 + 0.15i$ for $d = 100$ nm (dashed), 10 nm (solid) (see text) (f) Corresponding reflectance goes to zero for the 100 nm film and $\approx 0.1$ for the 10 nm film at the respective angles where the phasor sum approaches the origin.

**Figure 1**a shows the schematic of a thin film perfect absorber, with light incident from air ($n_1 = 1$) onto a thin film of refractive index $n_2 + ik_2$ of thickness $d$, on an opaque substrate ($n_3 + ik_3$). The reflectance of the system $R = |r|^2$ can be written as;[50]

$$r = \frac{r_{12} + r_{23}e^{2i\delta}}{1 + r_{12}r_{23}e^{2i\delta}} = \sum_{m=0}^{\infty} r_m \qquad (1)$$

where $r_{ij}$ is the Fresnel reflection coefficient for incidence from medium $i$ to $j$, $\delta = k_{z2}d$ and $k_{z2} = k_0\sqrt{\epsilon_2 - \epsilon_1 \sin^2\theta}$ is the wave vector component normal to the interface in the thin film, $\epsilon_i$ is the complex permittivity, $k_0$ is the free space wave number and $\theta$ is the angle of incidence from air. The reflection coefficient $r$ may also be expressed as the coherent sum of infinitely many partial waves $r_m$ (Figure 1a). $r_0$ corresponds to the partially reflected wave from the first interface ($= r_{12}$) and $r_m = (1 - r_{12}^2)r_{23}{}^m r_{21}{}^{(m-1)} e^{2mi\delta}$ for $m \geq 1$ correspond to those emerging from the cavity after $m$ roundtrips. The partial waves may be conveniently represented as phasors in the complex plane [Re($r$), Im($r$)].[8,9] With negligible losses, the interface phase shifts (i.e. on reflection or transmission) are typically either 0 or $\pi$, and resonant interference conditions are determined by the phase accumulated during propagation. For example, the simplest dielectric anti-reflection coating has a thickness of $d = \lambda/4n_2$ that provides the necessary $\pi$ propagation phase for destructive interference of reflected waves.[51] Such a quarter-wave thick, lossy film placed on an opaque substrate absorbs incident light completely since both reflection and transmission are simultaneously inhibited. The dashed arrows in Figure 1b show the phasors for such a PEC-backed conventional ($n_2 < 1$) quarter-wave thick absorber.[8] At normal incidence, the first reflection $r_0$ has a phase shift of $\pi$ (rarer to denser medium), which is cancelled out by the remaining partial waves that acquire $\pi$ phase from each reflection at the PEC substrate and roundtrip in the film. The corresponding phasor trajectory then ends up at the origin, giving $R = |r|^2 = 0$.

To overcome this thickness limit of PA, the primary dependence on film thickness for phase accumulation must be overcome, requiring interface phase shifts that deviate significantly from the lossless case to satisfy the resonance conditions. Such interferences exploiting non-trivial interface phase shifts have been demonstrated in ultrathin films by introducing weakly reflecting metal substrates,[8,9] metamaterial mirrors[52] or perfect magnetic mirrors[18] instead of PECs. If a weak reflector such as Au in visible frequencies is combined with a lossy thin film, the phasors are no longer horizontal (Figure 1c). In this case, the reflection phase at the *film-substrate* interface differs significantly from π, allowing the partial waves to cancel out $r_0$ even when $d \ll \lambda$. Solid arrows in Figure 1b shows the phasor trajectory of such a system [9] with parameters $d$ = 10 nm, $n_2$ = 4.3+0.7i and $n_3$ = 0.44+2.24i at λ = 532 nm, which returns to the origin tracing out a loop, corresponding to $d \approx \lambda/12n_2$.

However, constraints on the substrate material limits the applicability of these interferences on highly reflecting surfaces, [12,17] e.g. ubiquitous low cost metals such as Al and conventional metals such as Au, Ag at infrared wavelengths that behave increasingly like PECs. In order to achieve PA in ultrathin films backed by PECs (which impart fixed π phase), it is necessary to obtain non-trivial phase shifts at the *top* interface i.e. the air-film interface. For normal incidence at a lossless top interface with $n_1 < n_2$, e.g. air-glass, $r_{12}$ is a negative real number determined by the optical constants with a fixed phase of π. Even for highly absorbing dielectrics where $r_{12}$ becomes complex, the phase deviates by less than ~ $10^0$ from the lossless case. [10,53] However, if $n_1 > n_2$, light can undergo total internal reflection at oblique incidence. In such cases, $r_{12}$ becomes complex above the critical angle $\theta_c = \sin^{-1} n_2/n_1$, signifying the presence of evanescent waves in the rarer medium, and above $\theta_c$, incurs an angle-dependent reflection phase varying between 0 and π.[50] In most applications where $n_1$ = 1, this cannot be realized with conventional dielectrics. However, lossless ENZ or near-zero-index (NZI) materials[22,54] with refractive indices well below 1 (Figure 1d) would allow TER of light at their surfaces above their critical angles. [47,49] Importantly, non-trivial phase shifts are also available in lossy systems ($n_2+ik_2$) where the external reflection is not total, above an equivalent critical angle defined for the corresponding lossless material, i.e. $k_2$ = 0, $\theta_c = \sin^{-1} n_2/n_1$ (Supporting Information Section S1). This is demonstrated by the phasor diagrams in Figure 1e for *p*-polarized light of λ = 532 nm incident on a PEC-backed low-index layer with $n_2+ik_2$ = 0.45+0.15i. For d = 100 nm ($\approx \lambda/12n_2$), the phasors return back to the origin at an incident angle $\theta = 32^0$ (>$\theta_c = 27^0$). If $d = 10$ nm ($\approx \lambda/120n_2$), the reflectance reaches a minimum $\approx 0.1$ at $\theta = 68^0$. The corresponding numerically evaluated total reflectance for both thicknesses are plotted against $\theta$ in Figure 1f, showing their minima at the angles where the phasor sum reaches closest to the origin. Notably, in contrast to the highly absorbing 10 nm layer in Figure 1b, it takes many more roundtrips for the partial waves to converge to the final reflectance for the low-loss 10 nm layer in Figure 1e, stemming from the low loss nature of absorption that will be discussed later.

## 2.2. Spectral and angular characteristics of infrared absorption

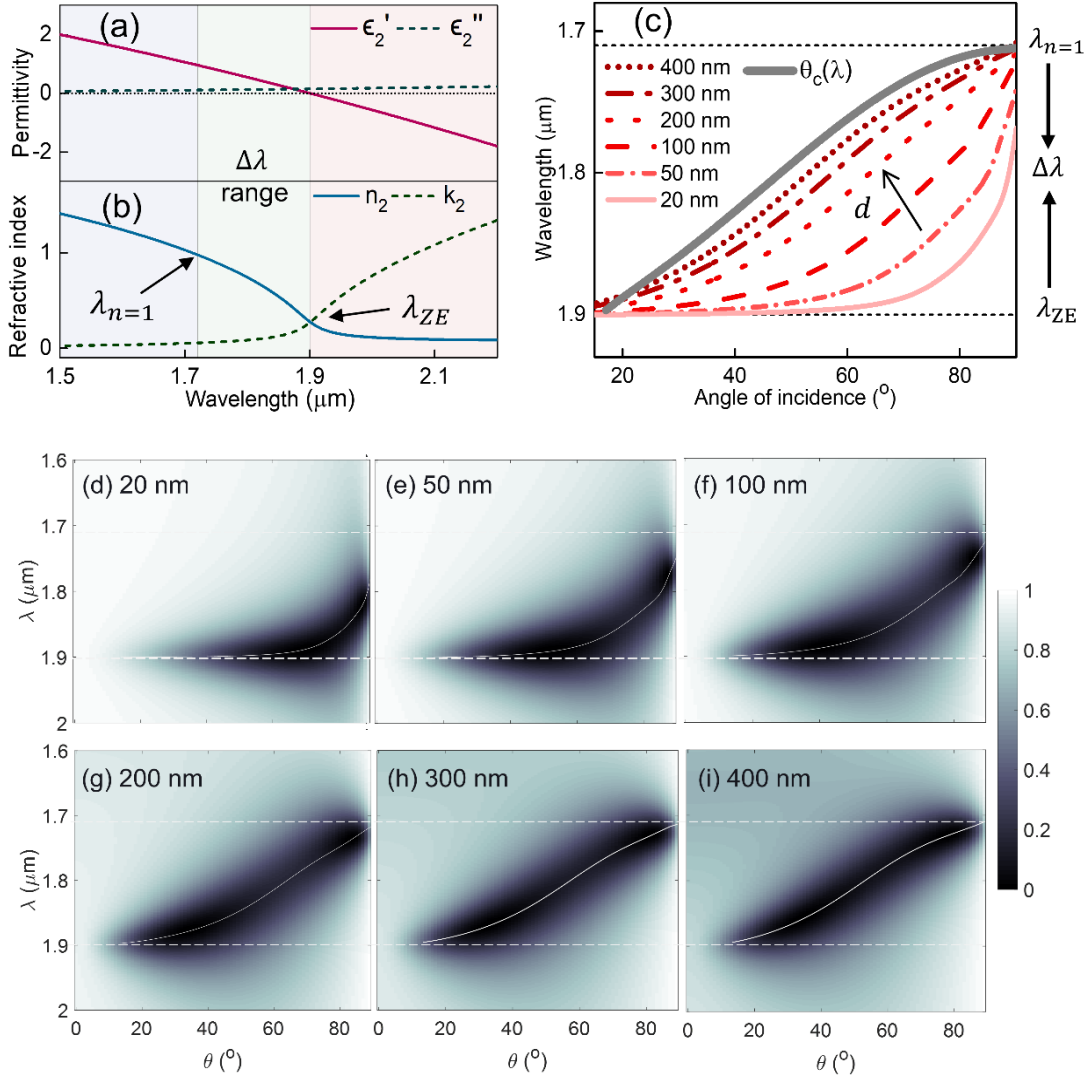

**Figure 2**. (a) Real ($\epsilon_2'$) and imaginary ($\epsilon_2''$) parts of ENZ medium permittivity. (b) Corresponding real ($n_2$) and imaginary ($k_2$) parts of refractive index, $n_2 + ik_2 = \sqrt{\epsilon_2}$. Vertical lines demarcate the $\Delta\lambda$ range between $\lambda_{n=1}$ and $\lambda_{ZE}$ where $n_2 < 1$ and $\epsilon_2' > 0$. (c) Dispersion relations of TM modes in CdO:Au structure for *d* varying from 20 to 400 nm, with the arrow showing direction of increasing *d*. The modes are confined to angles above the air-ENZ equivalent critical angle $\theta_c$ (solid grey curve) within the $\Delta\lambda$ range. (d-i) Reflectance maps of the CdO:Au structure in the $\lambda$ - $\theta$ plane for *p*-polarized light showing agreement with the dispersion relations in (c) overlaid as solid curves here. Horizontal dashed lines in (c) and (d-i) demarcate the $\Delta\lambda$ range.

To illustrate the spectral and angular characteristics of infrared absorption in continuous, planar media, we numerically analyze two low loss, ENZ film on metal (ENZ:M) structures. The permittivities of the ENZ medium ($\epsilon_2(\omega)$) are described by those of CdO in the near-IR[55] and SiO$_2$ in the mid-IR[56], and that of the substrate ($\epsilon_3(\omega)$) by Au, as outlined in SI Section S2. For the model parameters of CdO, the 'zero-epsilon wavelength' ($\lambda_{ZE}$) where $\epsilon_2' = 0$, lies at $\lambda_{ZE}$ = 1900 nm (**Figure 2**a). Figure 2b shows that $n_2 \approx 0.3$ at $\lambda_{ZE}$ and increases towards shorter $\lambda$, reaching 1 at $\lambda_{n=1} \approx 1720$ nm. In the wavelength window $\Delta\lambda$ between $\lambda_{n=1}$ and $\lambda_{ZE}$, the thin film is a dielectric with $n_2 < 1$,[57] where ER with the non-trivial phase shifts are realized. Figure 2c plots the dispersion relations for transverse magnetic modes in the CdO:Au system for $d$ varying from 20 to 400 nm (see SI Section S3 for details of numerical calculations). Interestingly, the dispersions are limited on the short wavelength side by $\lambda_{n=1}$, and on the long wavelength side by $\lambda_{ZE}$, above which the layer becomes metallic in nature i.e. $\epsilon_2' < 0$.[43,46,58] Figure 2d–i show reflectance maps of the CdO:Au system for *p*-polarized incident light as a function of $\theta$ and $\lambda$, evidencing well-defined darker regions of strong absorption (low reflectance), which closely follow the corresponding dispersion (white solid curves) and are confined to the $\Delta\lambda$ range. For comparison, Figure S2 in SI plots the reflectance spectra for a CdO:PEC system demonstrating a close match with the data above, indicating that Au is well approximated by a PEC in the investigated wavelength range.[59] The flat dispersion of the film with the lowest thickness (20 nm) is the well-known Berreman ENZ mode seen in very thin films nearly independent of its optical environment.[46,60] In thicker films, the ENZ mode dispersion moves away from $\lambda_{ZE}$ and the optical environment (symmetric or asymmetric, metallic or dielectric) may be expected to play a more prominent role in deciding the nature of supported modes. Since the highest thickness is 400 nm ($< \lambda/4$), the modes at higher thicknesses evidenced above are not conventional Fabry-Perot modes, which only start appearing for shorter wavelengths as shown in SI Figure S3. Furthermore, the wavelength-dependent critical angle, $\sin\theta_c(\lambda) = n_2(\lambda)$ (grey solid line in Figure 2c), overlaid on the dispersion relations shows that the dispersions lie to its right irrespective of the film thickness, i.e. the modes are supported only in the regime $\theta > \theta_c$, where light undergoes external reflection at the ENZ surface. The dispersion profiles gradually shift towards $\theta_c(\lambda)$ with increasing thickness, with the thickest layer dispersion nearly lying on $\theta_c(\lambda)$ (see Figure S4 for plots showing this for $d >$ 400 nm and for a detailed discussion on this limiting behavior). The critical angle approaches 90° as $n_2$ approaches 1 (at $\lambda_{n=1}$) and the radiative modes vanish at larger values of refractive index for $\lambda < \lambda_{n=1}$.

The corresponding plots for the $SiO_2$:Au system in the mid-IR are shown in **Figure 3**. Here, $n_2$ goes below 1 when $\lambda > 7.25$ $\mu m$ and the presence of the reststrahlen band above 8 $\mu m$ induces the ENZ condition.[58] Consequently, in the range $\Delta\lambda = 7.25 - 8$ $\mu m$, $SiO_2$ possesses a low-loss, below unity refractive index (Figure 3a). Figure 3(b-d) shows the reflectance maps of the system for $d = 100$, 1000 and 4000 nm. The lowest thickness shows a Berreman mode at $\lambda_{ZE} = 8$ $\mu m$ in Figure 3b. With increasing thickness, the absorption in the dielectric range moves away from the ENZ condition towards shorter wavelengths (Figure 3c). At higher thicknesses, it is evident that the absorption features lie over the dispersion of the critical angle (red curve in Figure 3b-d), restricted to the $\Delta\lambda$ range. Further reflection plots of the SiO2:Au system are presented in Figure S7 in SI.

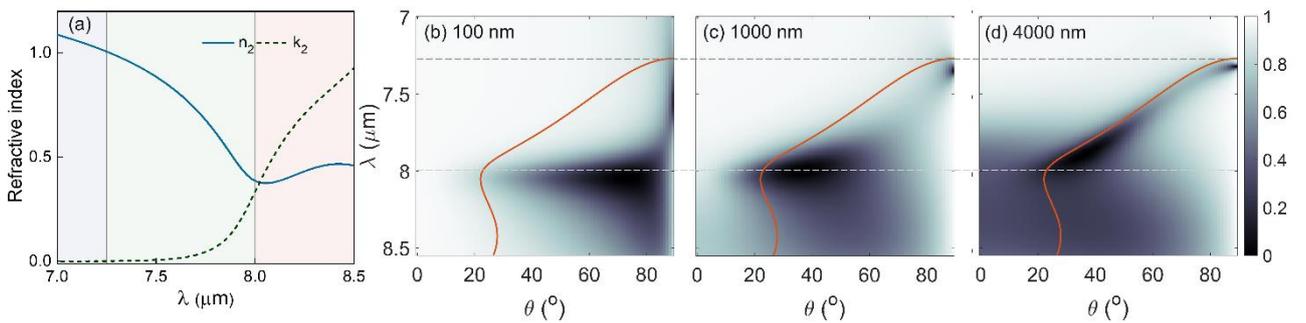

**Figure 3**. (a) Refractive index of $SiO_2$ demarcating a $\Delta\lambda$ range between 7.25 – 8 $\mu m$ (green shaded region). (b-d) Reflectance maps of $SiO_2$:Au system for p-polarized light for $d = 100$, 1000 and 4000 nm. Horizontal dashed lines demarcate the $\Delta\lambda$ range. Solid curve plots the critical angle $sin^{-1} n_2/n_1$ in all three panels and clearly demarcates the mode evolution towards $\theta_c$ and confinement to $n_2 < 1$ region.

It is worth noting that the gradual shift in dispersion with thickness and the limiting behavior at higher thicknesses where the dispersions approach $\theta_c(\lambda)$ are present in both ENZ systems, within their $n < 1$ spectral regimes. For the CdO ENZ system, this spectral range spans $\approx 200$ nm in the vicinity of $\lambda_{ZE}$ whereas for $SiO_2$, it is $\approx 1\mu$m. In both systems, the angular range $\theta > \theta_c$ is determined by the ENZ material dispersion, $\sin\theta_c(\lambda) = n_2(\lambda)$. To verify that these features arise from a general wave interference phenomenon not necessarily restricted to ENZ or polaritonic media, the reflectance of a system where the ENZ layer is replaced by a lossy dielectric with $n_2 = 1$ and the ambient medium has a refractive index greater than the dielectric layer is presented in SI Section S5. Strong absorption above the critical angle similar to those shown in Figure 2 and 3 are observed in this system, indicating that the absorption effects are a general feature of systems having the appropriate index contrast for external reflection. These results also justify the use of the critical angle calculated using the equivalent lossless media in demarcating the absorption regime in the investigated systems.

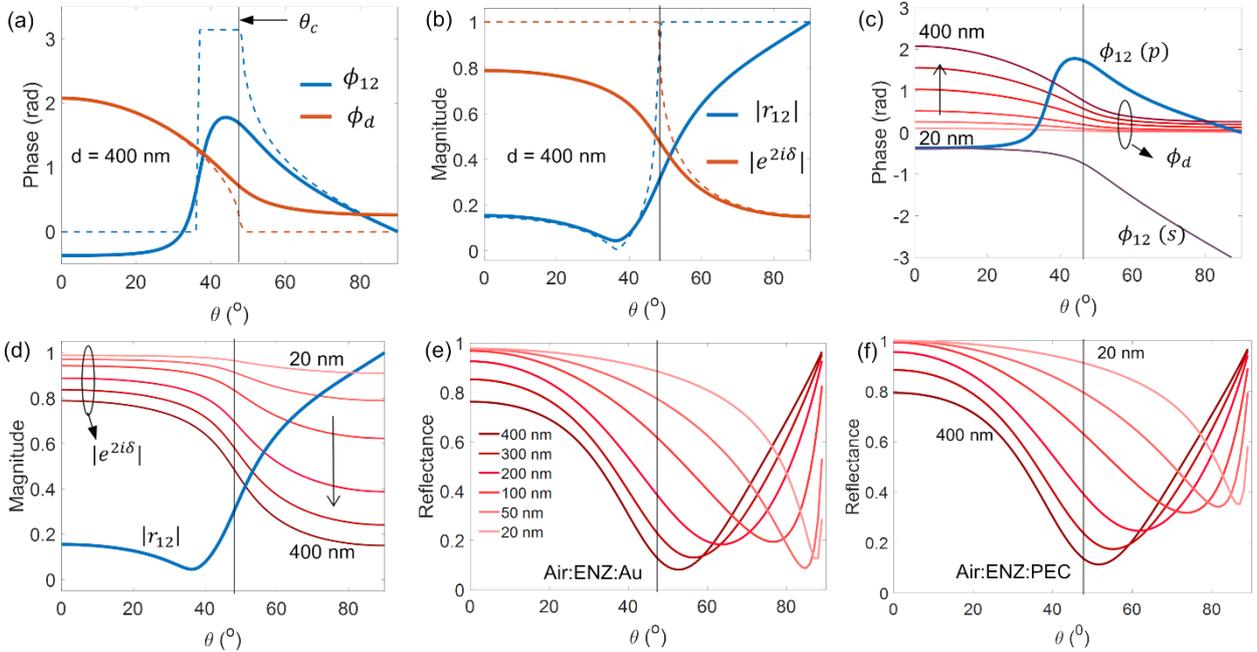

**Figure 4**: (a) Reflection phase shift $\phi_{12}$ (blue lines), and propagation phase $\phi_d$ (orange lines) versus $\theta$. (b) Magnitude of reflection at Air – ENZ interface (blue) and attenuation over a round trip in the film (orange) vs $\theta$. Plots in (a) and (b) are calculated for p-polarized light and $d = 400$ nm. Dashed lines show the lossless ENZ case where the effects of total external reflection are pronounced. (c) $\phi_{12}$ for s and p polarizations and $\phi_d$ for varying $d$ showing that the phases

match only for *p*-polarized light. (d) |*r₁₂*| and |$e^{2i\delta}$| for varying *d*. The exponential dependence of attenuation on *d* determines the resonant angles where the magnitudes match. (e,f) Reflectance vs *θ* for Au and PEC substrates showing the thickness dependent variation of reflection minima above *θ_c*. All calculations are done at λ = 1800 nm. The plots in (e) correspond to horizontal line cuts of the data in Figure 2(d-i).

To understand these universal features in the absorption of thin films in their $n < 1$ regime, we write the three medium reflection coefficient in **Equation 1** in the form $r = \frac{r_{12} - e^{2i\delta}}{1 - r_{12} e^{2i\delta}}$ by setting $r_{23} = -1$, corresponding to a PEC substrate.[50] Here, $e^{2i\delta}$ gives the phase advance (and attenuation) for a round trip in the film. For strong absorption, we require that the terms in the numerator of *r* cancel each other, giving two conditions to be satisfied simultaneously viz. $|r_{12}| \approx |e^{2i\delta}|$ and a phase difference $\approx 0$.[12] Physically, this signifies complete destructive interference between the reflected wave at the first interface and the resultant of the partial waves emerging from the thin film. **Figure 4**a plots the phases of *r₁₂* ($\phi_{12}$, reflection phase shift at the air-ENZ interface) and $e^{2i\delta}$ ($\phi_d$, propagation phase in the film) for CdO:Au with *d* = 400 nm, at a representative λ = 1800 nm lying in the *Δλ* range (solid lines). A comparison of these plots at three different wavelengths is shown in Figure S9. For the equivalent lossless ENZ (dashed lines), $\phi_{12}$ varies continuously in the range 0 to $\pi$ above *θ_c*. On the other hand, $\phi_d = 2 \operatorname{Re}(k_{z2})d$ is non-zero at low angles (< $\pi$ as $d < \lambda/4n_2$), and falls to *zero* above *θ_c*. Introducing loss in the film perturbs the phases as shown by solid lines in Figure 4a. The plot indicates that the phase shifts determining resonant interferences originate dominantly from the reflection phase at the top interface, in contrast to those in Fabry-Perot resonances (determined by propagation phase in the film) and the previously investigated ultrathin film interferences[11-20] (reflection phase at the bottom interface). Figure 4b plots the variation in magnitudes of *r₁₂* and $e^{2i\delta}$ for real and lossless ENZ. In the absence of loss (dashed lines), | *r₁₂*| rises up to unity at $\theta_c$ while $|e^{2i\delta}| = e^{-2 \operatorname{Im}(k_{z2})d}$ has unit magnitude below *θ_c* and drops off sharply above it as $\operatorname{Im}(k_{z2})$ becomes non-zero. Both magnitudes are limited by loss as shown by solid lines when $k_{z2}$ becomes complex valued. Evidently, the crossover between the two magnitudes occurs near $\theta_c$ for both lossless and lossy media. Note that these features are generally valid at wavelengths where external reflection occurs i.e. in the $\Delta\lambda$ window. Together, Figure 4a,b show that $r$ will be minimum near $\theta_c$ corresponding to an optimal match between the magnitudes and phases for $d = 400$ nm, explaining why its dispersion in Figure 2c lies close to $\theta_c(\lambda)$ throughout the $\Delta\lambda$ range.

In order to explain the thickness and polarization dependence of the dispersions, $\phi_{12}$ for both $s$ and $p$ polarizations at $\lambda = 1800$ nm are plotted along with $\phi_d$ for $d$ in the range 20 – 400 nm in Figure 4c. Evidently, $\phi_{12}$ for $s$-polarized light cannot match $\phi_d$ for any thickness in the investigated range, explaining why interferences effects are observed only for $p$-polarization. It is worth noting that using the appropriate Fresnel reflection coefficients for magnetic media, similar effects may be expected in the case of materials having a near zero magnetic susceptibility. In contrast to its phase, $|e^{2i\delta}|$ in Figure 4d has a pronounced thickness dependence originating from its exponential dependence on $d$. In general, the phases match in two angle ranges: one at low angles and one at larger angles. Importantly, $|e^{2i\delta}|$ and $|r_{12}|$ match only at angles above $\theta_c$, due to the steep drop off in $|e^{2i\delta}|$ above the critical angle as the wave vector in the medium becomes dominantly imaginary and the increase in $|r_{12}|$ due to the effect of external reflection. Further, the two magnitudes match at larger angles for lower thicknesses. Together, these observations qualitatively explain the confinement to angles above $\theta_c$ and the shift in reflectance minima with thickness in Figure 2 and 3, exemplified in Figure 4e,f by the reflectance plots for the ENZ layer on Au and PEC substrates at $\lambda = 1800$ nm. Thus, thickness can act as a useful design parameter for tailoring dispersion in ENZ media, in addition to the well explored carrier density tunability of the ENZ spectral range.[57]

## 2.3. Tailoring thermal emission

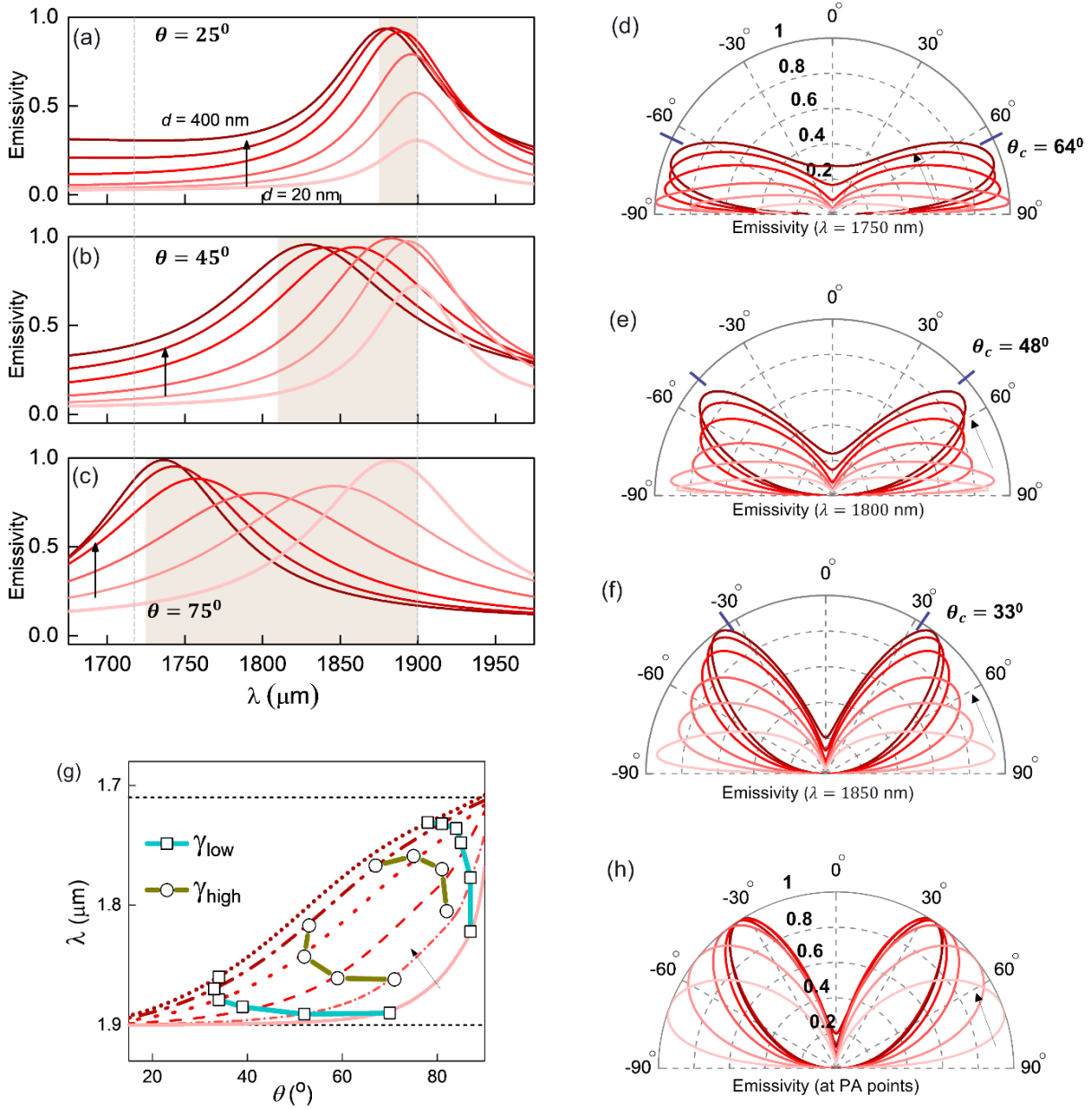

Figure 5: (a-c) Emissivity spectrum of CdO:Au structure at (a) $\theta = 25^0$, (b) $45^0$ and (c) $75^0$. The shaded regions in each panel shows the allowed spectral range of dispersions i.e. wavelengths lying below the dispersion of the critical angle in Figure 2c. Dashed vertical lines demarcate the $\Delta\lambda$ range. (d-f) Polar plots of emissivity vs angle of incidence of the same structure at (d) $\lambda = 1750$ nm, (e) 1800 nm and (f) 1850 nm, showing the variation of angular spectrum above the respective critical angles with ENZ thickness. The critical angle is marked on the angle axis. (g) Loci of perfect absorption points in the $\lambda$ - $\theta$ plane for various ENZ thicknesses (20 to 400 nm) and two values of loss parameter $\gamma$ in the near-IR system. Co-plotted are the thickness dependent dispersions in Figure 2c corresponding to the $\gamma_{low}$ case. (h) Emissivity at the respective PA points of each ENZ thickness from $d = 20 - 400$ nm for the low loss case ($\gamma_{low}$) corresponding to the low angle PA points lying near $\lambda_{ZE}$ (squares in (e)). An arrow shows the direction of increasing $d$ in each panel (a-f).

To illustrate the potential application of these results in tailoring thermal emission, we analyze the emissivity of the CdO:Au structure in its $\Delta\lambda$ range, calculated as 1-$R$, where $R$ is the reflectance of the structure.[35] **Figure 5**a-c shows the emissivity as a function of wavelength for varying CdO thickness ($d$ = 20 – 400 nm) at three angles: $25^0$, $45^0$ and $75^0$. At low angles, the peaks lie close to $\lambda_{ZE}$ for all thicknesses and the spectral variation of emissivity with $d$ is negligible, whereas the emissivity peak becomes increasingly thickness dependent at larger angles. This behavior can be understood from Figure 2c, where the modes at any given angle are constrained to lie at wavelengths below the dispersion of $\theta_c(\lambda)$. At low angles, the spectral range available for the modes is limited close to the vicinity of $\lambda_{ZE}$, constraining the emissivity peaks close to $\lambda_{ZE}$ as demarcated by the shaded region in Figure 5a. At larger angles, as the trajectory of $\theta_c(\lambda)$ moves away from the vicinity of $\lambda_{ZE}$, the modes can be supported over a larger range of wavelengths resulting in a wider emissivity range, indicated by the wider shaded region encompassing the peaks in Figure 5b. At very large angles, the available wavelength range approaches the whole of the $\Delta\lambda$ range (dashed vertical lines in Figure 5a-c), allowing a large variation in the spectral emissivity with $d$. This is commensurate with the behavior of $\theta_c(\lambda)$ that approaches $\lambda_{n=1}$ at very large angles, allowing mode dispersions to extend over the whole $\Delta\lambda$ range.

In order to explore the angular dependence of emissivity of the ENZ structure, polar plots in Figure 5(d-f) compare emissivity as a function of angle of incidence at three different wavelengths in the $\Delta\lambda$ range of CdO: 1750, 1800 and 1850 nm. A notable feature at each of these wavelengths is the thickness dependent angular selectivity of the emissivity, which can be tailored from grazing angles for the lowest thicknesses up to the respective $\theta_c$ for larger thicknesses. For example, the emissivity is restricted to lie above $\theta_c = 64^0$ at 1750 nm and $\theta_c = 33^0$ at 1850 nm for all the ENZ thicknesses, calculated from the refractive index of the ENZ medium. Such control over angular emission can be useful in thermal camouflaging and directional beaming applications.[35] The angle-dependent features of thermal emissivity can also be straightforwardly understood from the dispersion relations in Figure 2c, where at any given wavelength in the $\Delta\lambda$ range, the modes are constrained to lie towards the right of $\theta_c(\lambda)$, i.e. at angles above the critical angle.

Finally, we analyze the locations of zero reflectance, i.e. PA, along the dispersions and its dependence on material loss. (see SI Section S7 for numerical methods). Figure 5g shows the PA locations for varying $d$ (20 – 400 nm) and for two values of loss (scattering rate): $\gamma_{low}$, typical value for doped CdO (squares) and $\gamma_{high}$, typical value for indium tin oxide (circles) (see SI Section S2 for material parameters). In general, there are two PA points for a given value of loss and thickness, one at lower angles and another at higher angles.[43] The low angle PA points lie closer to $\lambda_{ZE}$ for lower losses. Figure 5h shows the corresponding emissivities calculated for the $\gamma_{low}$ system for the various thicknesses at their respective PA wavelengths that vary between 1850 – 1900 nm (low angle PA). Unit emissivities, corresponding to perfect absorption of light, are observed with high angular and spectral selectivity depending on wavelength and ENZ thickness, illustrating highly efficient tailored thermal emission in these sub-wavelength ENZ layers. Although the tunability and limiting ranges of ENZ thermal emissivity are demonstrated here at near-IR wavelengths, the results presented earlier for the case of $SiO_2$, a mid-IR ENZ material, shows that these results are equally applicable in the infrared regions of interest for thermal emission such as the atmospheric transparency window (8-13 μm).

A combination of low loss and low thickness in a perfect absorber may be counter-intuitive considering that a reduced absorber volume is expected to be compensated by sufficiently high loss. However, as remarked earlier, low loss PA in metal-backed ENZ layers has been predicted, which requires lower loss for *thinner* films.[42,61] To explain this, we note that the power dissipated per unit volume in an absorbing medium is given by $P = \epsilon_0 \epsilon'' \omega |E|^2 / 2$, where $\epsilon_0$ is the free space permittivity, $\epsilon''$ is the imaginary part of permittivity and $E$ is the electric field.

Clearly, the effect of low loss ($\epsilon''$) may be compensated by increased electric field strengths. Considering a PEC-backed low loss ENZ layer (Figure 1d) that satisfies the PA condition for *p*-polarized light, the absence of a reflected field implies that the incident and transmitted fields are related by the continuity relation $E_{z1} = \epsilon_2 E_{z2}$ and the field enhancement in the ENZ medium is $|E_{z2}|/|E_{z1}| \approx 1/\epsilon_2''$. Thus, the field in the ENZ layer scales as $|E_2| \propto 1/\epsilon_2''$. This gives $P \propto 1/\epsilon_2''$, leading to stronger absorption for lower loss in ENZ systems. Additionally, this explains why the PA points for the low loss case lie closer to $\lambda_{ZE}$ than the high loss case. The material loss plays an important role in determining infrared absorption and emission linewidths as well as the achievable field enhancements in the structures. The electric field profiles and enhancement factors are compared for the two losses and varying thicknesses at and away from PA conditions in SI Section S8. The high field enhancement at PA may be effectively exploited for a variety of applications e.g. designing high efficiency photodetectors.[25,62]

## 3. Conclusion

In conclusion, we have shown that resonant interferences on a highly reflecting surface can be sustained employing an ultrathin film of *n* < 1, surpassing the quarter-wave thickness limit for perfect absorption on a PEC surface. The continuous range of phase shifts available on external reflection at the air-film interface provides the necessary phases to satisfy the destructive interference conditions. This is in contrast with conventional interferences that depend on propagation phase or previously investigated ultrathin film interferences exploiting substrate reflection phases. The established conditions for ultrathin film interferences such as weakly reflecting substrates and/or large intrinsic material loss are overcome here, allowing absorption resonances even on PEC substrates, albeit at oblique incidence. The low loss realization, a unique feature distinct from previous demonstrations that used highly lossy films or substrate materials, is achieved only in the ENZ limit due to the enhanced ENZ field-assisted dissipation. The spectral and angular ranges of the resonance and its dependence on the ENZ thickness, polarization and loss are well explained by simple analytical expressions of the reflection coefficient of the structure, enabling precise tailoring of the mode dispersions for engineering absorption and thermal emission characteristics in the infrared. The spectral and angular ranges within which the dispersions can exist ($\Delta\lambda$ range and $\theta > \theta_c$) are shown to be the limits within which emissivity can be tailored in ENZ systems. The potential of the tailored dispersions in exerting precise control over both the spectral and angular thermal emissivity of the ENZ-metal bilayer is explored, showing unique abilities to control both the emissive spectral and angular ranges as well as the ability to fine-tune emissivity by tuning the thickness of the ENZ layer. Although our results are complementary to the Berreman mode description of absorption in ENZ ultrathin layers, the wave interference picture captures the behavior of a wider class of materials that notably do not possess either ENZ or polariton resonances. The results described here provide fundamental insights into thin film optics of low index media such as ENZ and near-zero-index media and should extend the scope of applicability of ultrathin film resonant phenomena. Juxtaposed with the readily tailored optical properties of ENZ materials via chemical doping, optical pumping or electrical gating that tunes their $\Delta\lambda$ range, our results readily open up possibilities for designing ultrathin infrared optical absorbers and thermal emitters.

**Supporting Information**

Supporting Information is attached at the end of this document.

**Conflict of interest**

The authors declare no conflict of interest.

**Acknowledgements**

B.J. developed the concept and performed the calculations. All authors contributed to analyzing and interpreting data and writing the manuscript. The authors thank Dr. Ravi Pant (IISER-Thiruvananthapuram), Dr. Vijith Kalathingal (NUS) and Prof. Paul Dawson (QUB) for helpful discussions. The authors acknowledge financial support from SERB, Govt. of India (SR/52/CMP-0139/2012, CRG/2019/004965), UGC-UKIERI 184-16/2017(IC) and the Royal Academy of Engineering, Newton Bhabha Fund, UK (IAPPI_77). B.J. and S.C. acknowledge research fellowship from IISER Thiruvananthapuram.

# Supporting Information

**Tailoring Infrared Absorption and Thermal Emission with Ultrathin-film Interferences in Epsilon-Near-Zero Media**

*Ben Johns\*, Shashwata Chattopadhyay and Joy Mitra\**
School of Physics, Indian Institute of Science Education and Research, Thiruvananthapuram, India 695551

E-mail: ben16@iisertvm.ac.in, j.mitra@iisertvm.ac.in

**Section S1.  Analyzing reflection coefficient in the case of external reflection**

The polarization-dependent Fresnel reflection coefficients $r_{ij}$ at an interface of media $i$ and $j$ are given by[1]

$$r_{ij} = \frac{\epsilon_i k_{zj} - \epsilon_j k_{zi}}{\epsilon_i k_{zj} + \epsilon_j k_{zi}} \quad (p\text{-polarization}) \tag{S1}$$

$$r_{ij} = \frac{k_{zi} - k_{zj}}{k_{zi} + k_{zj}} \quad (s\text{-polarization}) \tag{S2}$$

Where $k_{zi}$ (or $k_{zj}$) is the normal wave vector component in medium $i$ (or $j$). For zero total reflectance, eqn. 1 in main text requires that two relations must be satisfied simultaneously:

$$\phi_{12} - (\phi_{23} + \phi_d) = (2m+1)\pi \tag{S3}$$

$$|r_{12}| = |r_{23} e^{2i\delta}| = |r_{23}| e^{-2d\,Im(k_{z2})} \tag{S4}$$

where $r_{ij} = |r_{ij}| e^{i\phi_{ij}}$, $\phi_{ij}$ are reflection phase shifts, $\phi_d = 2d\,Re(k_{z2})$ is the propagation phase accumulated by a roundtrip in the film and $m$ is an integer. A reflectance minimum is observed when these two equations are approximately satisfied.

For a PEC substrate, $r_{23} = -1$. This simplifies the above two equations to:

$$\phi_{12} = \phi_d + 2m\pi \tag{S5}$$

$$|r_{12}| = |e^{2i\delta}| = e^{-2d\,Im(k_{z2})} \tag{S6}$$

For the case of $n_2 < 1$ such as an ENZ or NZI medium, in the lossless case the reflection phase shifts are angle dependent above the critical angle [Fig. S1(a), dashed curves], given by:

$$\phi_{12(p)} = 2\tan^{-1}\frac{n\sqrt{n^2 \sin^2\theta - 1}}{\cos\theta} \quad (\phi_{12},\ p\text{-polarization}) \tag{S7}$$

$$\phi_{12(s)} = 2\tan^{-1}\frac{\sqrt{n^2 \sin^2\theta - 1}}{n\cos\theta} \quad (\phi_{12},\ s\text{-polarization}) \tag{S8}$$

where $n = \frac{n_1}{n_2}$. In the presence of loss $(n_2 + ik_2)$, analogous phase shifts are still obtained as seen from solid curves in Fig. S1(a). Fig. S1(b) plots the magnitude of $r_{12}$ showing the effect of loss on total external reflection. In the lossy case, this is referred to as external reflection due to the fact that reflection is not total anymore. The validity of TER in the limit of low losses can be inferred by plotting the magnitude of the reflection coefficient for decreasing values of loss, as shown in the figure S1(c), plotted for $\gamma$ varying from 0.018 eV to 0.0 eV. In the low loss limit, the external reflection tends to become total. The ENZ material here is doped CdO.

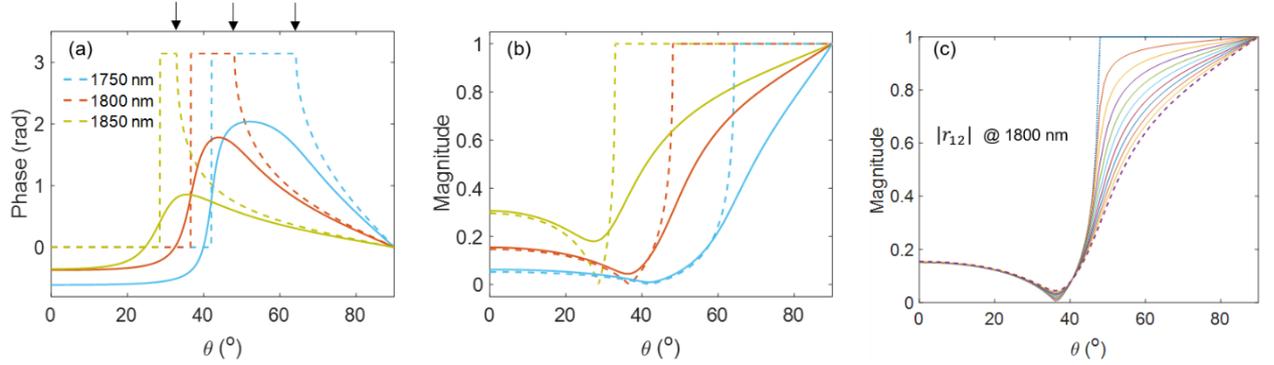

Figure S6: (a) Reflection phase $\phi_{12}$ and (b) magnitude $|r_{12}|$ versus angle of incidence for $p$-polarized light incident at the air-ENZ interface, for $n_2+ik_2 = 0.9+0.06i$, $0.75+0.08i$ and $0.54+0.13i$, corresponding to critical angles $\theta_c = 64^0, 48^0$ and $33^0$ marked by vertical arrows. The dashed lines show the lossless phases and magnitudes in (a) and (b), setting $k_2 = 0$. The continuous variation of $\phi_{12}$ above critical angle in (a) is described by eqn. S7 in the lossless case. Solid lines show calculations with loss that are numerically evaluated from eqn. S1. In (b), the sharp variation in $|r_{12}|$ at angles near the critical angle is clear in the lossless case while the presence of loss prevents complete (unity) reflection above the critical angle (solid lines). (c) $|r_{12}|$ vs $\theta$ at $\lambda = 1800$ nm for loss varied from 0 to the actual value for CdO $\gamma_{low}$, in steps of $\gamma_{low}/10$.

**Defining the critical angle**

The critical angle is defined as $\sin \theta_c = n_2/n_1$, where $n_2 + ik_2 = \sqrt{\epsilon'_2 + i\epsilon''_2}$. Although total external reflection is not well defined for a lossy medium, the $\theta_c$ is calculated for a lossless material equivalent with the same real refractive index to demarcate the external reflection region of interest that contribute to the observed absorption features. The use of this nominal critical angle in the case of lossy media is motivated by the following reasons:

1. The reflection minima in Fig 2 and 3 clearly reach a limiting case with increasing ENZ layer thickness (ref $\theta_c(\lambda)$ curve in Fig 2c and 3d). The discussion on Fig 4 and Fig S4-S6 provide further evidence that the limiting case is the $\theta_c$ defined for the lossless limit.

2. The significance of $\theta_c$ however is more general, which is demonstrated in Fig S8, where resonant interferences are demonstrated in dielectric structures without ENZ or polaritonic resonances. The numerical results show that absorption features restricted to lie above the nominal critical angle at the top interface are found in these structures as well.

These observations guide us to the conclusion that the absorption in the lossy system is confined to angles above $\theta_c$, defined for the lossless system.

**Section S2.    Thin film and substrate permittivities**

Dielectric function of the ENZ layer is given by a modified Drude permittivity model,

$$\epsilon_2(\omega) = \epsilon_\infty - \omega_p^2/(\omega^2 + i\omega\gamma) = \epsilon_2' + i\epsilon_2'' \qquad (S9)$$

where $\omega$ is the angular frequency, $\epsilon_\infty = 5.3$ is the background (high frequency) permittivity, $\omega_p = 2.28 \times 10^{15}$ rad/s (1.5 eV) is the plasma frequency, $\lambda_{ZE} = 1900$ nm (zero-epsilon wavelength), and $\gamma = 2.8 \times 10^{13}$ rad/s (0.018 eV) is the scattering rate, corresponding to typical values for doped CdO in this wavelength range [2].

The substrate gold (Au) is described by a Drude model approximation, $\epsilon_3(\omega) = 1 - \omega_{p,Au}^2/(\omega^2 + i\omega\gamma_{Au})$ with $\omega_{p,Au} = 1.367 \times 10^{16}$ rad/s (9 eV) and $\gamma_{Au} = 1.52 \times 10^{14}$ rad/s (0.1 eV) [3].

For the comparison of two different losses in Fig. 5 of main text, the values used are $\gamma_{low} = 2.8 \times 10^{13}$ rad/s (0.018 eV) and $\gamma_{high} = 6 \times 10^{13}$ rad/s (0.04 eV).

$\gamma_{low}$ corresponds to typical value for CdO in this wavelength range [2].

$\gamma_{high}$ corresponds to typical value for indium tin oxide (ITO) in this wavelength range [4].

**Section S3.    Eigenmode analysis**

The dispersion relation for TM modes in a three medium structure is obtained as a solution $(k_x, \omega)$ of the equation [5]

$$1 + \frac{\epsilon_1 k_{z3}}{\epsilon_3 k_{z1}} = i\tan(k_{z2}d)\left(\frac{\epsilon_2 k_{z3}}{\epsilon_3 k_{z2}} + \frac{\epsilon_1 k_{z2}}{\epsilon_2 k_{z1}}\right) \qquad (S10)$$

where $k_x$ is the in-plane wave number and $k_{zi}$ the transverse wave number in medium $i = 1,2,3$, which are related by $k_{zi}^2 = \epsilon_i \omega^2/c^2 - k_x^2$. A real $k_x$ and complex $\omega$ representation is chosen here, in accordance with previous numerical studies on such systems [6,7]. Fig 2(c) in the main text is plotted by converting the $\omega - k$ relations to the $\lambda - \theta$ plane using the relations $\omega = 2\pi c/\lambda$ and $k_x = (2\pi/\lambda)\sin\theta$.

## Section S4. Supplementary reflectance plots

### A. Reflectance with a PEC substrate

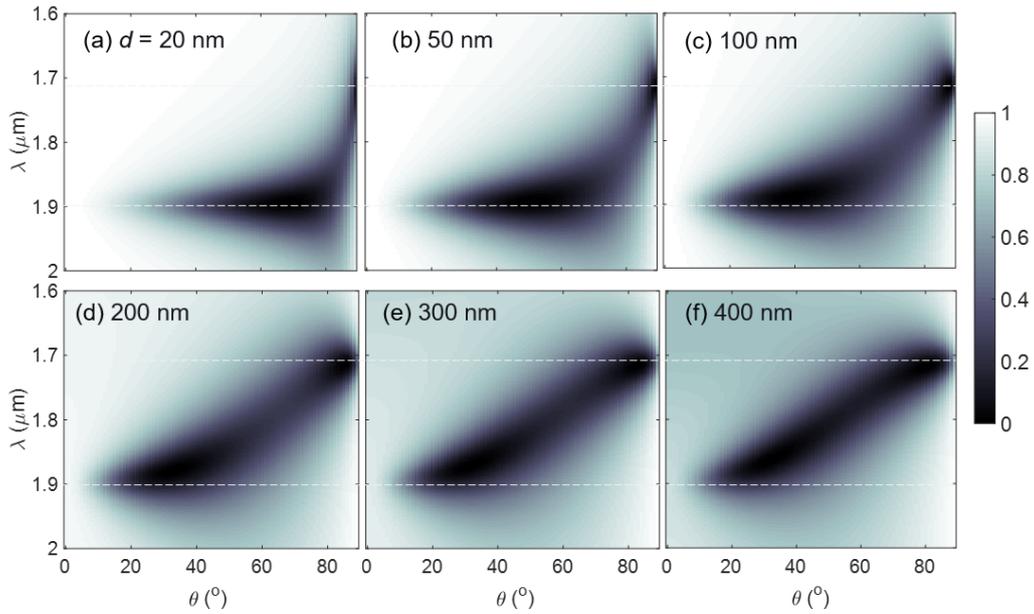

Figure S7: Reflectance of doped CdO films of varying thickness on a perfect electric conductor (PEC), i.e. $r_{23} = -1$, showing absorption features similar to that obtained with an Au substrate in Fig 2 in main text.

### B. Fabry-Perot modes in doped CdO on Au at lower wavelengths

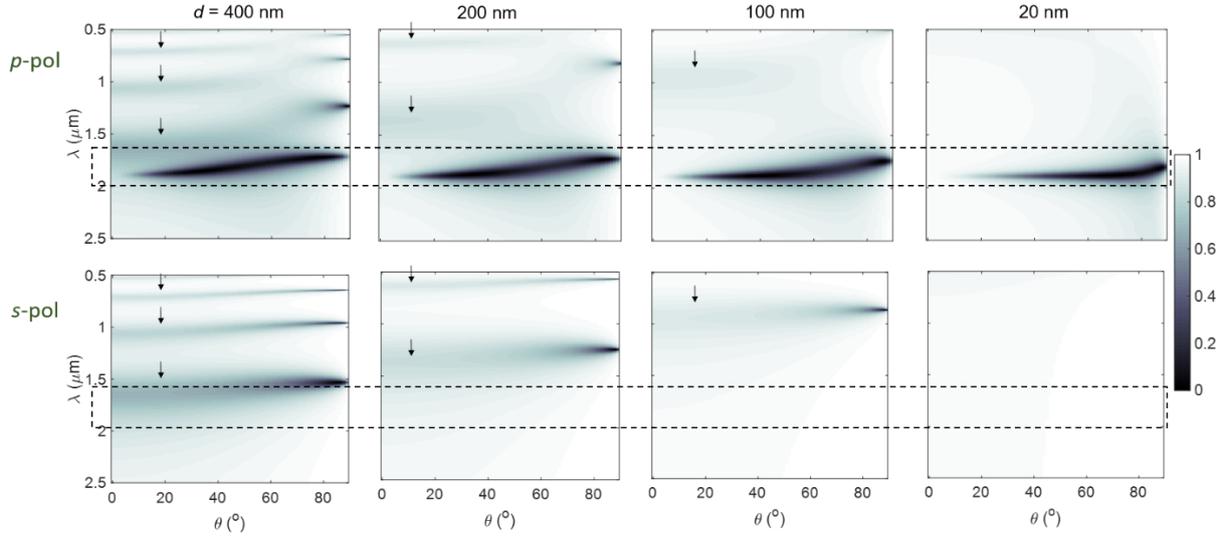

Figure S8: Reflectance maps in Fig 2 of main text are shown here for an extended wavelength range from 500 – 2500 nm for *p*- (top) and *s*-polarization (bottom panels). Quarter-wave and higher order Fabry-Perot modes are indicated by arrows. For the thickest film ($d$ = 400 nm), conventional Fabry-Perot modes are seen at low wavelengths for both polarizations (dark regions). With decreasing thickness from left to right, the modes move to lower wavelengths and finally disappear, leaving only the absorption of *p*-polarized light in the ENZ regime. The boxes indicate the range of wavelengths presented in the main text.

## C. Reflectance and dispersion of thicker films: Limiting behavior

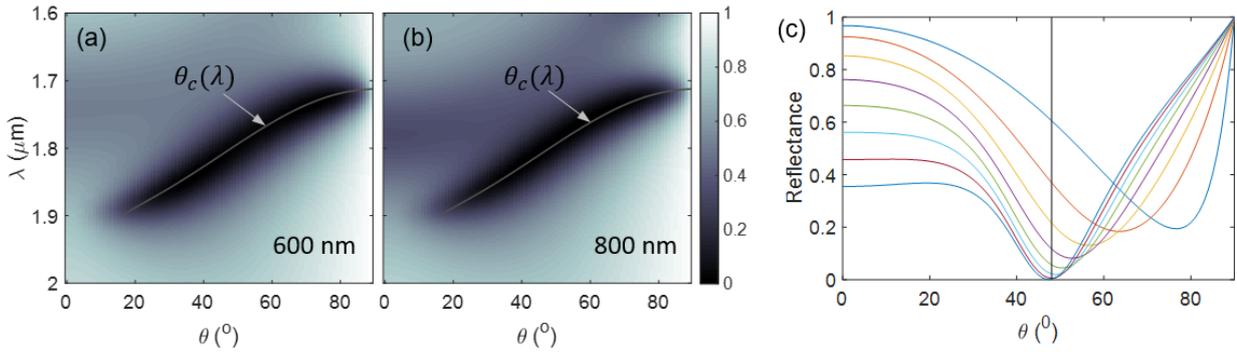

Figure S9: Fig 2 in main text plots reflectance of CdO:Au up to $d$ = 400 nm. Reflectance for thicker films is shown in (a) $d$ = 600 and (b) 800 nm, clearly revealing that the absorption profile is limited by the critical angle at higher thicknesses (solid line). Fabry-Perot modes also start appearing within the $\Delta\lambda$ range at these thicknesses, as evidenced by darker areas towards the short wavelength side. (c) For further evidence, the reflectance at $\lambda$ = 1800 nm as a function of incidence angle is plotted for thicknesses 100, 200, ... upto 800 nm, showing that the minima tend to lie on $\theta_c$ (vertical line) at higher thicknesses.

To investigate whether the limiting behavior of the dispersions at higher thicknesses is due to the ENZ layer acting as a bulk medium, we calculate the penetration depth of light into the ENZ layer to ascertain if only the ENZ surface is relevant at larger thicknesses. Fig S5 plots the penetration depth of light into the ENZ layer, $1/Im(k_{z2})$, at three different wavelengths within the range studied in this work. In order to compare with the CdO ENZ layer thickness, the penetration depths at the reflection minima angle for the thickest layer ($d = 400$ nm) are marked by vertical lines. The penetration depth of light into the ENZ medium is $> 800$ nm, much larger than the film thicknesses considered here. This indicates that the limiting behaviour is not due to the ENZ layer acting as a semi-infinite medium at higher thicknesses. The light reflected from the substrate cannot be neglected while calculating the overall reflectance of the structure at these thicknesses and the effect of multiple wave reflection within the ENZ layer should be considered for an accurate analysis of the reflectance.

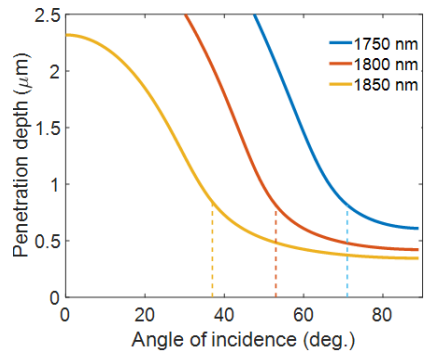

Figure S10: Penetration depth vs angle of incidence at three different wavelengths in CdO ENZ medium

In ENZ media, the Brewster's angle is approximately equal to the critical angle [8]. To verify that the limiting behavior is not a Brewster angle absorption, Fig S6 compares the dispersion of the Brewster's angle (dashed line) and critical angle (solid line) with the absorption at the largest investigated thicknesses: 400 nm CdO (Fig S6a) and 5000 nm $SiO_2$ (Fig S6b) on Au. Evidently, the absorption approaches the critical angle, and not the Brewster's angle, in the limiting case. The Brewster's angle is calculated numerically as the angle of minimum reflectance at the air-ENZ interface to take into account the ENZ loss.

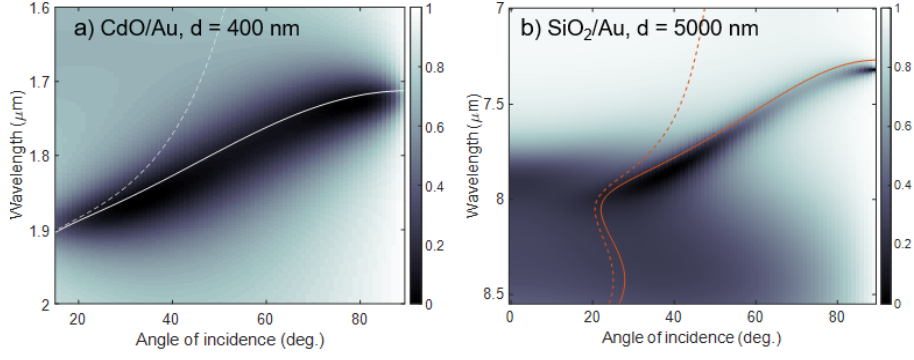

Figure S11: Comparison of the dispersion of the Brewster's angle in lossy medium (dashed) and critical angle (solid line) with the absorption at relatively large ENZ thickness for the two ENZ:M systems.

### D. Reflectance of SiO2:Au system

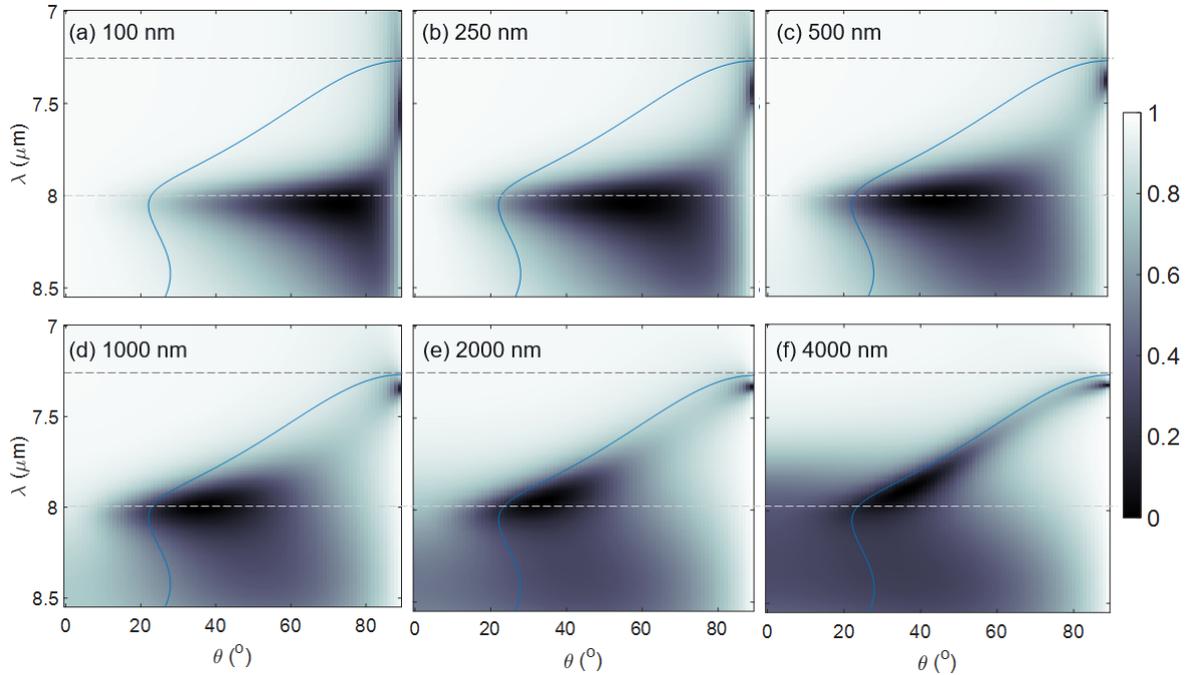

Figure S12: Reflectance maps of SiO$_2$:Au for *p*-polarized light showing gradual evolution of modes with thickness towards the critical angle (solid blue curve). Horizontal dashed lines demarcate the $\Delta\lambda$ range where $n_2 < 1$.

### Section S5.  Interference in the absence of ENZ or polaritonic media

As evident from the reflectance maps in Fig. 2 and 3 of the main text for CdO:Au and SiO$_2$:Au structures, the thinnest films with $d \sim \lambda/50$ supports a Berreman mode (radiative ENZ mode) excitation, with a predominantly flat absorption profile at $\lambda_{ZE}$. The dispersions progressively blue-shift with increasing film thickness and have a qualitatively different shape for larger thicknesses, becoming nearly identical as they are limited by the wavelength-dependent critical angle. To investigate if these results are independent of polaritonic or ENZ effects and depend only on external reflection at the top interface, we calculated the reflectance of an analogous system. In this system, the ENZ layer is replaced with a lossy dielectric having $n_2 = 1$ and the top dielectric medium is chosen to have $n_1 > 1$, providing the required index contrast at the top interface while precluding ENZ effects in the thin film layer. The top panel in Fig. S8 shows for comparison the calculated reflectance vs $\theta$ at three wavelengths in the $\Delta\lambda$ range (1750, 1800 and 1850 nm) for various thicknesses of the CdO ENZ layer film on a PEC substrate, corresponding to horizontal line cuts of the plots in Fig. S2. Evidently, reflection minima appear only above $\theta_c$, as marked by dashed vertical lines. Now, in the second system with a lossy dielectric between a higher index dielectric and PEC substrate, $n_1$ is chosen such that the critical angle at the top interface, $\theta_c = \sin^{-1} n_2/n_1$, remains the same as in the ENZ on PEC case, i.e. $n_2/n_1$ is the same (Fig S8, bottom panel). For example, at $\lambda = 1800$ nm, the real part of ENZ refractive index is $n_2 = 0.74$, corresponding to $\theta_c = \sin^{-1} 0.74/1 = 48^0$. In the corresponding second system, the $n_1$ is chosen such that $\theta_c = \sin^{-1} 1/n_1 = 48^0$, giving $n_1 = 1.34$. Indeed, the reflection minima again appear above the critical angle in the bottom panels, showing similar absorption as in the ENZ system. To include loss, an imaginary refractive index of $k_2 = 0.2$ is considered for the dielectric layer in all calculations. Note that although the second system is analogous to the Otto configuration, the absorption does not arise from surface plasmon excitation as the PEC substrate does not support any surface waves.

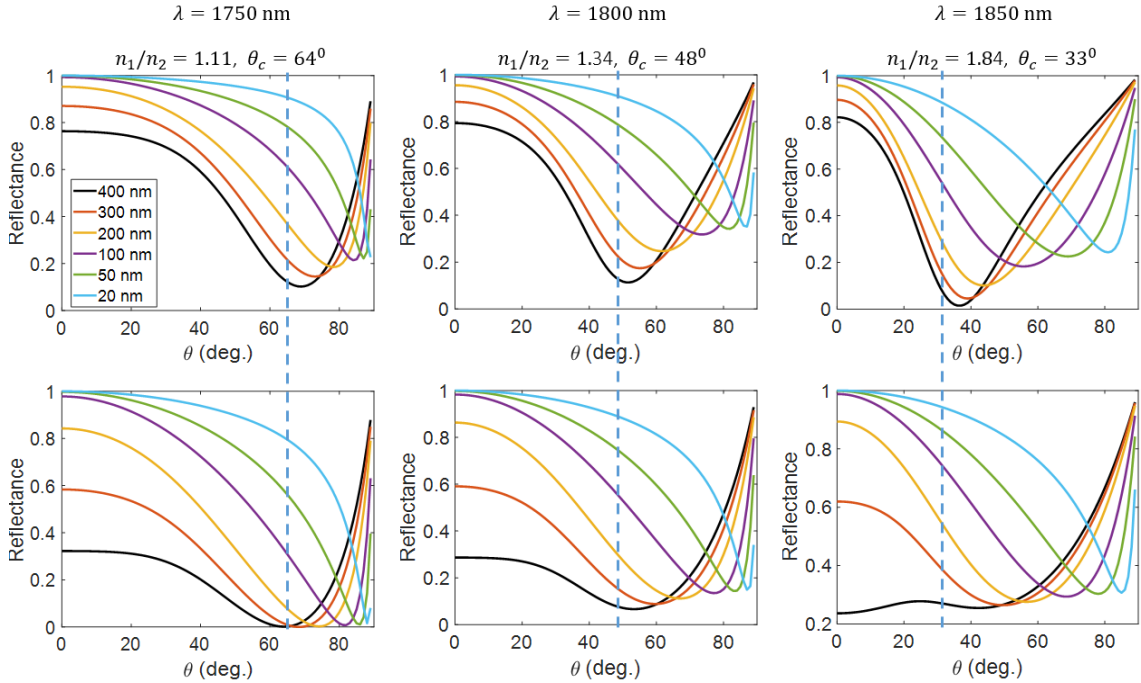

Figure S13: Calculated reflectance vs angle of incidence for Air:ENZ:PEC (top) and high index dielectric: low index dielectric: PEC (bottom). The wavelengths, refractive indices and critical angles are noted in text inset. Dashed vertical lines denote the critical angle, above which reflection minima are seen in both systems.

## Section S6.  Comparison of phase and magnitude of $r_{12}$ and $e^{2i\delta}$

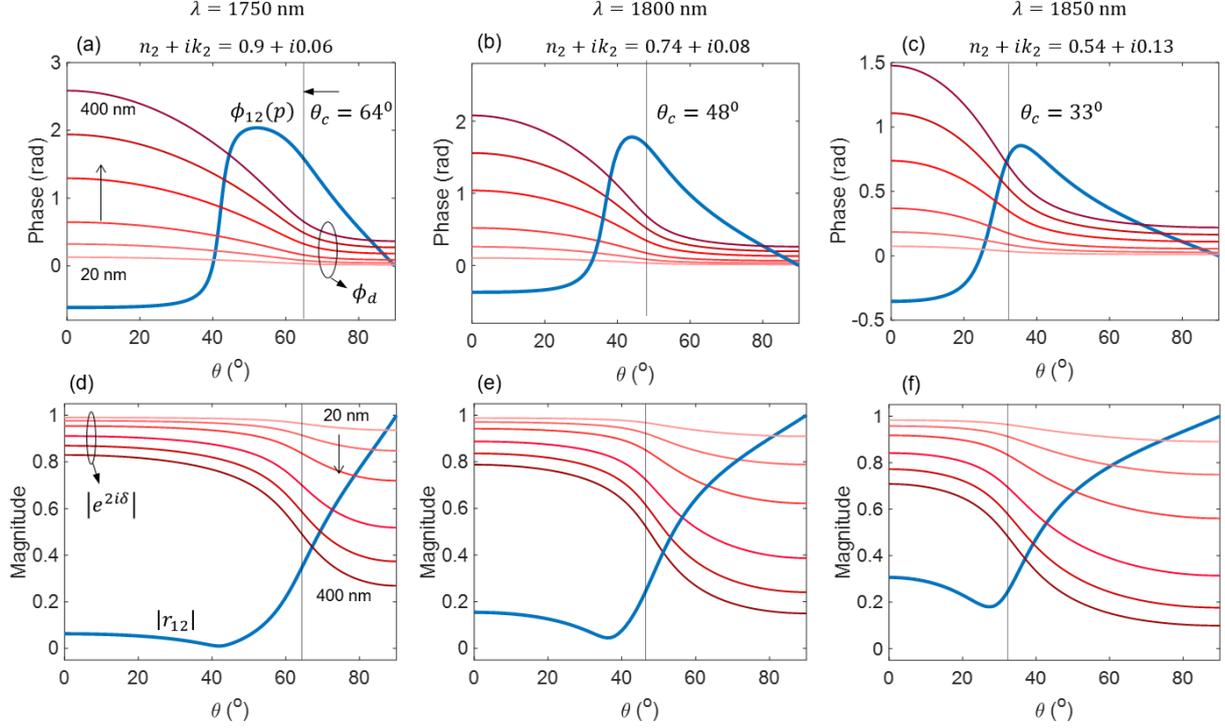

Figure S14: Comparison of phases (a-c) and magnitudes (d-f) of $r_{12}$ and $e^{2i\delta}$ for the ENZ layer, showing the intersection points where eqns. S5 and S6 are satisfied. From left to right, vertical panels correspond to wavelengths $\lambda = 1750$ nm, 1800 nm and 1850 nm respectively. In general, the phases match in two angle ranges: one below the critical angle and one at larger angles. However, the magnitudes match only above the critical angle (vertical lines) as seen in the bottom panel, restricting the absorption to angles above $\theta_c$.

## Section S7.  Analysis of perfect absorption

While various complementary descriptions may be used to analyze perfect absorption in the system, we use the formalism of Luk et al [6] to identify PA locations from the eigenmode analysis described in SI section 3. From the real wave number–complex frequency solutions of the dispersion relation in eqn. S10, the condition for PA is a zero imaginary part of $\omega$, i.e. $Im(\omega) = 0$. A positive (negative) imaginary part of the mode frequency corresponds to growing (decaying) fields in the structure. The sign of $k_{z1}$ is chosen such that it always corresponds to an incoming wave in medium 1 (air). A zero imaginary part thus corresponds to an incoming plane wave in the first medium that is completely absorbed in the structure without any reflection. Figure S11a,b shows the trajectory of modes in the complex frequency plane i.e. $Im\ (\omega)$ vs $Re\ (\omega)$ for thicknesses from 20 nm to 400 nm for $\gamma_{low} = 2.8 \times 10^{13}$ and $\gamma_{high} = 6.0 \times 10^{13}$ rad/s respectively. The PA locations are identified where the trajectories cross the x-axis, which agree well with the reflectance minima extracted numerically from eqn. 1 in the main text.

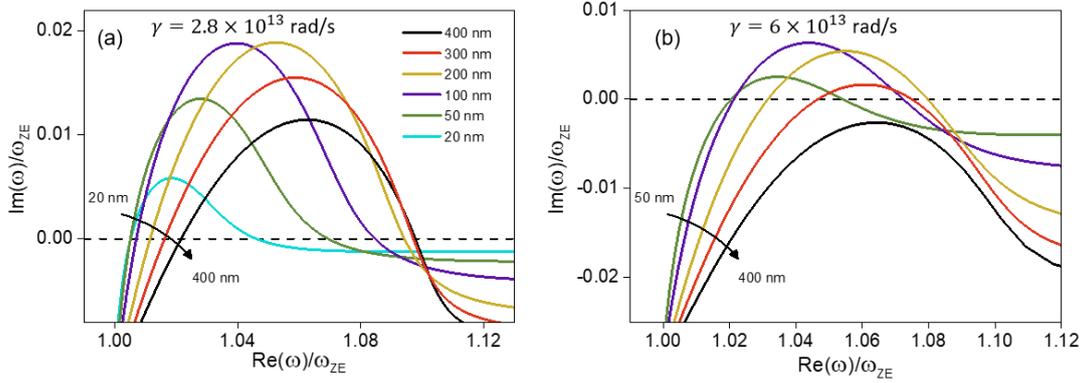

Figure S15: Plots of $Im\ (\omega)$ vs $Re\ (\omega)$ normalized to $\omega_{ZE}$ for (a) $\gamma_{low} = 2.8 \times 10^{13}$ rad/s and (b) $\gamma_{high} = 6 \times 10^{13}$ rad/s. In general, the trajectories cross the x-axis at two values of $Re\ (\omega)$, denoting two PA locations. To plot the corresponding PA points in the $\lambda - \theta$ plane shown in Fig. 5 of main text, the values of $Re\ (\omega)$ are used to identify the corresponding $\theta$ values from the respective dispersion relations. In (a), all the thicknesses have two PA locations while in (b), no PA condition is possible for the thickest film ($d$ = 400 nm), which does not cross the x-axis in this frequency range.

### Section S8. Field profiles at and away from the PA condition

Figure S12 (a-e) shows color maps of the local field enhancement for the *z*-component of the field ($E_z$) with respect to the incident (background) field, at the respective PA wavelengths for three CdO film thicknesses and two values of $\gamma$. Realization of PA is characterized by the absence of any reflected wave in air. The plots further show the variation of field enhancement across the layers (solid blue/cyan lines), quantifying the field enhancement within the ENZ layer. It is worth noting that for equal thickness, the film with higher $\gamma$ exhibits weaker field enhancement compared to that with lower $\gamma$. Fig S12(f) plots the percentage of light absorbed in the Au substrate as a function of ENZ film thickness. Interestingly, even for the low-loss thinnest film (20 nm), only 4% of the incident light is dissipated in the substrate with the remaining 96% absorbed by the thin film, which rises to 99% in the thicker films. Fig. S13 similarly plots the field enhancement for samples at $\lambda = 1750$ nm, away from their PA conditions. Here, the non-zero reflectance manifests itself in the field pattern in air and the subdued field enhancement in the ENZ medium.

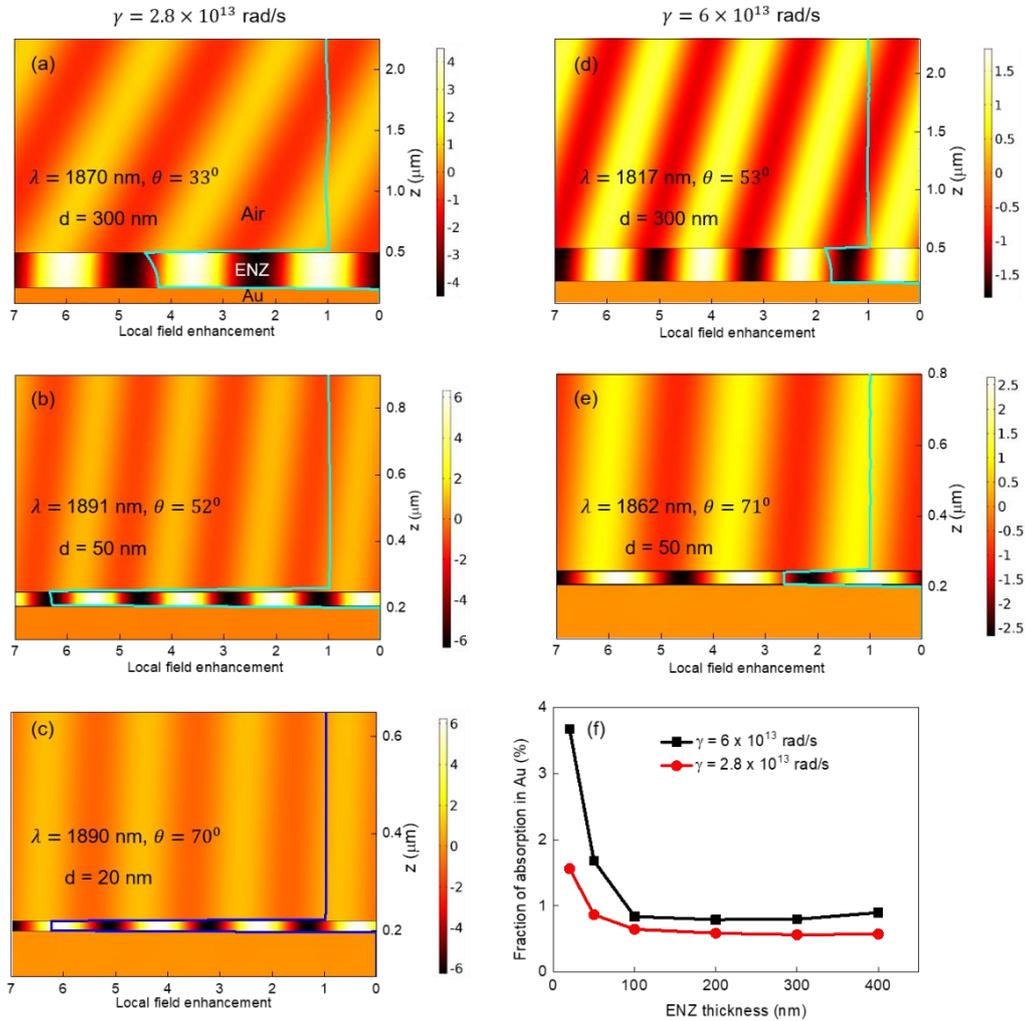

Figure S16: (a-e) Electric field distribution in the CdO:Au system for various ENZ film thickness and two losses at their respective PA conditions (see text inset and Fig. 5 in main text). Solid cyan/blue lines plot $|E_y|$ enhancement across the 3 layers, overlaid on its spatial profile (color map) showing local electric field strength in the three media. Spatial field variation in air is characterized by the absence of any outgoing reflected wave due to PA. Higher $|E_y|$ in the ENZ layer is achieved for the lower loss case, determined by the absolute value of permittivity at the PA wavelength. (f) Percentage of the energy absorbed in the Au substrate for various ENZ thickness for the two losses.

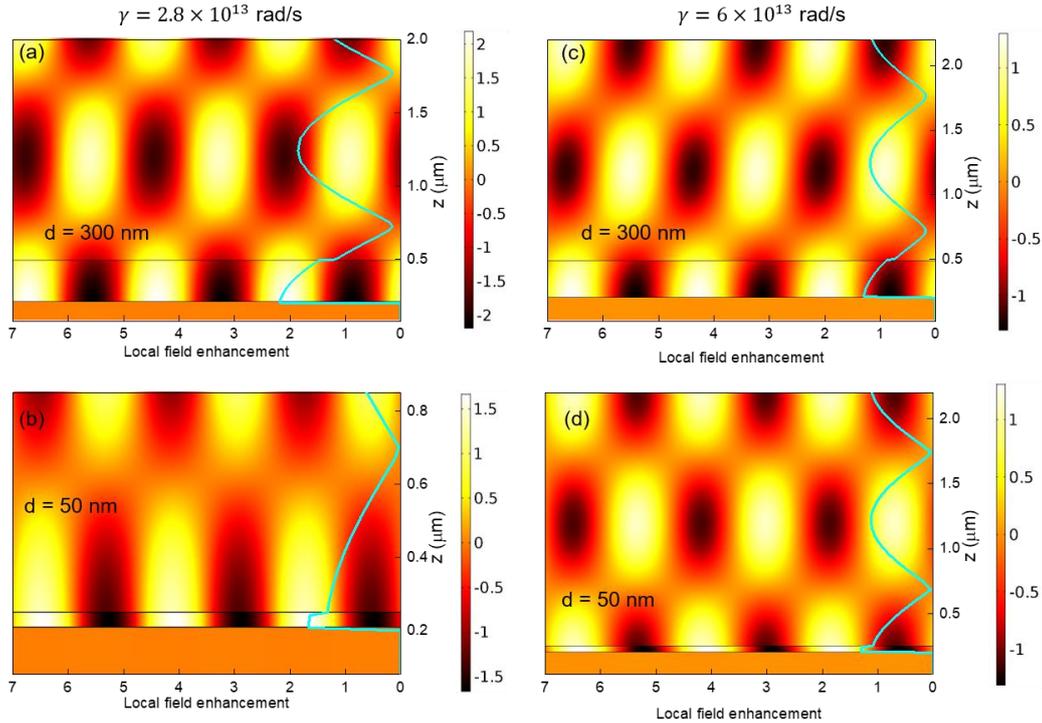

Figure S17: Variation in enhancement of $|E_y|$ across the 3 layers for $\lambda = 1750$ nm and $\theta = 33^0$ (away from PA conditions) for two different thickness and loss values. In contrast to Fig S12, field profiles in air clearly reveal superposition of counter-propagating waves (incident and reflected).